# Modeling dependence of creep recovery behavior on relaxation time distribution of ageing colloidal suspensions§

Yogesh M. Joshi

Department of Chemical Engineering,

Indian Institute of Technology – Kanpur,

Kanpur 208016, INDIA.

E-Mail: joshi@iitk.ac.in, Tel: 0091 512 259 7993, Fax: 0091 512 259 0104

**ABSTRACT**

A scaling model is developed to correlate relaxation time distribution of soft glassy materials to ultimate recovery. We propose that in the limit of creep-recovery time smaller than age, time translational invariance can be applied to ageing soft materials. In such limit, multimode linear viscoelastic model with Spriggs relaxation spectrum predicts enhancement in the ultimate recovery with broadening of the relaxation time distribution. We analyze these results in the context of creep-recovery behavior of aqueous suspension of laponite with varying concentration of salt.

## I. INTRODUCTION

Highly viscous, thixotropic pasty materials form the important constituents used in the chemical industry. These include concentrated suspensions,[1] concentrated emulsions and paints,[2] inks,[3] industrial slurries,[4] pharmaceutical and cosmetic creams, waxy derivatives of petroleum industry,[5] etc.[1, 6] By virtue of very high viscosity, most of the above mentioned materials do not reach equilibrium over the practical time scales. While understanding rheological behavior of these materials is very important from the processing point of view, a very complex microstructure and strong history dependence often pose significant problems in analyzing these materials.[7] In addition, time and deformation field dependent evolution of the microstructure often make it difficult to relate the rheological behavior to the micro-structural details. In this paper we have presented a qualitative model to relate creep-recovery behavior of thixotropic soft solids of aqueous suspension of laponite observed by Reddy and Joshi[8] to its relaxation time distribution.

Above mentioned highly viscous thixotropic soft materials are commonly known as soft glassy materials.[9] In these materials, translational diffusion above a certain length scale is highly constrained due to physical jamming of the structure at that length scale. This restricts the jammed entity to span only a part of the available phase space and hence fall out of equilibrium. Every material has a natural tendency to attain equilibrium.[10] For soft glassy materials, even though there may not be a global energy minimum, evolution of the microstructure progressively lowers the energy of various jammed particles with respect to time. This phenomenon is generally known as ageing. Although the ageing dynamics is strongly dependent on the microstructure of the material, every jammed particle, which has energetic interactions with its surrounding constituents, can be considered to be confined to a cage-like environment. The cumulative energetic interactions of an arrested particle can be represented by energy well. Usually there exists a distribution of the energy well depths,[11, 12] by virtue of which there exists a distribution of relaxation times. The ageing dynamics causes lowering of the well depths that leads to slowing down of the relaxation process with the age (waiting time) of the system.

Traditionally various optical techniques have been employed to characterize the soft materials.[13-33] The major advantage of the same is the information they give about the relevant time scales, its distribution and the dominating length scales in the material without physically disturbing the same.[13, 17, 26, 29] Rheological study is important to carry out effective processing of such materials. Rheological characterization also gives significant physical insight with regard to the structure of the soft materials.[8, 34-37] However, it is an intrusive technique and hence disturbs the material. Generally it is observed that structural evolution (aging) under flow is slower compared to that under "no flow" conditions.[38-41] The corresponding structural evolution as represented by the dependence of dominating relaxation time scale on age (waiting time) can be estimated by carrying out systematic rheological experiments at different ages and by shifting the data appropriately to yield a master curve.[8, 38, 39, 42, 43] However, compared to optical characterization, there are fewer reports in the literature that study the relevant time scales, its distribution and the dominating length scales in the soft glassy materials using rheological tools.

In this work, we have developed a simple scaling model to qualitatively correlate creep recovery behavior of soft glassy materials to their relaxation time distribution. We discuss the results in the context of creep recovery behavior of aqueous suspension of laponite with varying concentration of salt reported by our group.[8] In order to understand various issues associated with the physical and rheological behavior of aqueous suspension of laponite, we discuss the same briefly in the next section.

## II. SYSTEM

Laponite, smectite hectorite clay, is composed of disc shaped nano-particles having 25 nm diameter with 1 nm of layer thickness.[20] Due to isomorphic substitution within the layer, the face of a laponite disc is negatively charged. The edge of the laponite disc is composed of hydrous oxides and is positive or negative depending upon pH of the system.[44] At pH 10, an electrostatic screening length associated with the face is 30 nm,[15] however there is no consensus regarding the nature of the edge charge at this pH.[22, 45] The magnitude of the edge charge is weaker than that of a face. This leads to strong repulsion among the laponite particles causing ergodicity breaking at the low concentration of laponite.[15, 46] However, the possibility of the edge-to-face attractive interactions can not be ruled out in the overall repulsive environment.[21, 47] Addition of sodium chloride increases the concentration of $Na^+$ ions that screen the negative charges on the face enhancing attraction among the particles.[27, 28] It is generally observed that the microstructure gets progressively less homogeneous with addition of salt.[27-29, 47] Thus the microstructure gets strongly affected by the concentration of salt. The phase behavior of laponite suspension with respect to concentration of laponite and that of the salt is extensively debated in the literature, and has lead to different proposals for the phase bahavior.[17, 21, 22, 28, 33, 48]

Recently, in a series of papers, our group has published detailed study of the rheological behavior of aqueous suspension of laponite at varying concentration of salt.[8, 34, 38, 46, 49] It was observed that the elastic modulus of this system is a strong function of salt concentration and is known to increase with the waiting time (age).[34] Joshi and coworkers carried out systematic creep experiments at predetermined value of elastic modulus and waiting time.[8, 34] Although, the modulus at which the creep experiments were carried out was same; interestingly, the system with less concentration of salt showed more recovery [refer to figure 3 of Reddy and Joshi[8]]. This behavior was observed irrespective of the value of modulus. In this work we analyze this behavior by developing a model to qualitatively correlate structure and dynamics of aqueous suspension of laponite with varying concentration of salt.

### III. MODEL

Joshi and coworkers[34, 46] observed that soon after mixing laponite in ultra-pure water, system enters a non-ergodic state. All the samples on which creep-recovery experiments were carried out were confirmed to be in the non-ergodic state as elastic and viscous modulus showed weak dependence on frequency.[9, 34] We have earlier demonstrated that creep behavior of such systems can be represented by single mode linear Maxwell model (linear viscoelastic fluid).[34, 38] However, since single mode Maxwell model does not show the observed behavior (different recovery at the same elastic modulus), we need to consider the distribution of relaxation times.

Let us consider a multimode linear viscoelastic fluid undergoing creep deformation. It is known that ageing systems fail to obey time translational invariance (TTI) under certain circumstances, and hence we begin by considering a linear viscoelastic model without TTI given by:[9]

$$\boldsymbol{\sigma}(t) = \int_{-\infty}^{t} G(t-t_w, t_w) \dot{\boldsymbol{\gamma}}(t_w) dt_w \tag{1}$$

where $\boldsymbol{\sigma}(t)$ is the stress tensor, $\dot{\boldsymbol{\gamma}}$ is rate of strain tensor, and

$$G(t-t_w, t_w) = \sum_k \mathscr{G}_k [\xi_k(t) - \xi_k(t_w)] \tag{2}$$

is the relaxation modulus. It states that stress at present time $t$ not only depends on time elapsed between present time $t$ and past time $t_w$, the time at which deformation was applied; but also the state of material at time $t_w$. The latter contribution considers exclusive effect of ageing on the material behavior. Fielding et al.[9] proposed that general form of function $\xi(t)$ is given by,

$$\xi(t) = (t/\theta)^{1-\mu}, \tag{3}$$

where $\theta$ is the microscopic time and the function $\xi(t)$ interpolates between TTI for $\mu=0$ and simple ageing for $\mu=1$ [$\xi(t) \sim \ln t$]. It can be easily shown that in the limit $t - t_w \ll t_w$, response function can be written as:

$$\xi_k(t) - \xi_k(t_w) = \frac{1-\mu}{\theta^{1-\mu}} \frac{t-t_w}{t_w^\mu}. \tag{4}$$

If we represent $\mathscr{G} \sim \exp(-\xi)$, the stress relaxation modulus can then be written as:

$$G(t-t_w, t_w) = \sum_k G_k e^{-(t-t_w)/\tau_k}, \tag{5}$$

where $\tau_k \left( \sim \theta_k^{1-\mu_k} t_w^{\mu_k} \right)$ and $G_k$ represent spectrum of relaxation times and modulus respectively. Equation 5 represents a very interesting feature which shows that in the limit $t - t_w \ll t_w$, $t_w$ dependence of $G(t-t_w, t_w)$ solely comes from relaxation time dependence of age. Furthermore, since $t-t_w \ll t_w$, change in age over creep-recovery timescale can be neglected which makes $\tau_k$ to be constant over the application of creep. Under such conditions, we may apply TTI to the equation (1) and we will treat equation (1) as a linear Maxwell model that obeys Boltzmann's superposition principle and hence TTI. This analysis is in accordance with the analysis of short term creep proposed by Struik[43] which indeed apply TTI to the ageing polymeric glasses by making similar arguments.

We can represent $\tau_k$ such that $\tau_k > \tau_{k+1}$. The elastic and viscous moduli for linear viscoelastic model represented by equation 1 and equation 5 with TTI are given by:[50]

$$G' = \sum_k \frac{G_k \tau_k^2 \omega^2}{1+\tau_k^2 \omega^2} \text{ and } G'' = \sum_k \frac{G_k \tau_k \omega}{1+\tau_k^2 \omega^2}. \tag{6}$$

Ultimate recovery after creep for this model is given by,[50]

$$\gamma_\infty = \sigma_0 \int_0^\infty sG(s)ds \bigg/ \left[\int_0^\infty G(s)ds\right]^2, \tag{7}$$

where $\sigma_0$ is the value of the shear stress at which the creep experiment was carried out. In order to use above expression for ultimate recovery, the creep flow is needed to be carried out for long enough time to ensure the steady state flow. Experiments of Reddy and Joshi[8] fulfill this condition [refer to figure 3 of the same paper] such that the creep time was long enough to achieve steady state, but small enough to have total creep-recovery time sufficiently smaller than waiting time or age of the sample.

It can be seen from equation 7 that both the moduli and the ultimate recovery depend strongly on the relaxation spectrum. The problem can be significantly simplified by assuming a representative empirical relaxation spectrum suggested by Spriggs[51] as detailed in the literature:[50, 52]

$$\tau_k = \tau_0 / k^\alpha, \text{ and} \tag{8}$$

$$G_k = G_0 k^{-\alpha} \bigg/ \sum_k k^{-\alpha}, \tag{9}$$

where $\alpha$ is a dimensionless quantity and represents breadth of the relaxation spectrum. Equation 8 clearly suggests that larger value of $\alpha$ leads to narrower breadth of the relaxation spectrum. This empiricism is frequently used in modeling polymeric liquids with $\alpha$ in the range 2 to 4.[50] Combining equations 6 to 9 yield the following expressions for elastic modulus and ultimate recovery respectively:

$$G' = G_0 \left\{ \sum_k \frac{k^{-3\alpha}}{\tau_0^{-2}\omega^{-2} + k^{-2\alpha}} \right\} \bigg/ \left( \sum_k k^{-\alpha} \right), \tag{10}$$

$$\gamma_\infty = (\sigma_0/G_0) \left\{ \left( \sum_k k^{-\alpha} \right) \left( \sum_k k^{-3\alpha} \right) \bigg/ \left( \sum_k k^{-2\alpha} \right)^2 \right\}. \tag{11}$$

The values of elastic modulus at which Reddy and Joshi[8] have reported the creep experiments are 190 Pa and 314 Pa (they observed same behavior at larger values of elastic modulus as well). The longest relaxation time at this stage is significantly high ($\tau_0 >> 10$ s).[34] Since values of $\omega$ for rheological experiments are generally in the range $0.1 < \omega < 100$ (rad/s), it is safe to assume that $\tau_0\omega >> 1$. In addition, in a non-ergodic state, elastic modulus is weakly dependent on the frequency, with lowest accessible frequency limited by the age.[9] Under such condition equation 10 is further simplified to yield $G' \approx G_0$, irrespective of the value of $\alpha$. However, according to equation 11, the ultimate recovery still depends on the value of $\alpha$. This simple approach thus illustrates that samples exhibiting similar values of $G'$ (or $G_0$), show different ultimate recovery by virtue of different breadth of relaxation spectrum. Figure 1 shows normalized ultimate recovery $(\gamma_\infty G_0/\sigma_0)$ plotted against $\alpha$. It can be seen that for samples having identical creep stress, $\sigma_0$ and storage modulus, $G'$ (or $G_0$); ultimate recovery increases by broadening the breadth of the spectrum (i. e., decrease in $\alpha$).

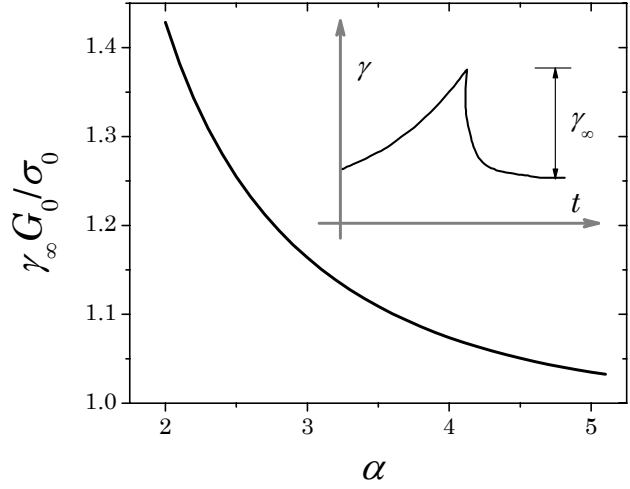

**Figure 1.** Normalized ultimate recovery is plotted against $\alpha$ given by equation 11. As $\alpha$ increases (or the breadth of relaxation spectrum decreases), ultimate recovery decreases. Inset shows cartoon of creep-recovery curves representing ultimate recovery.

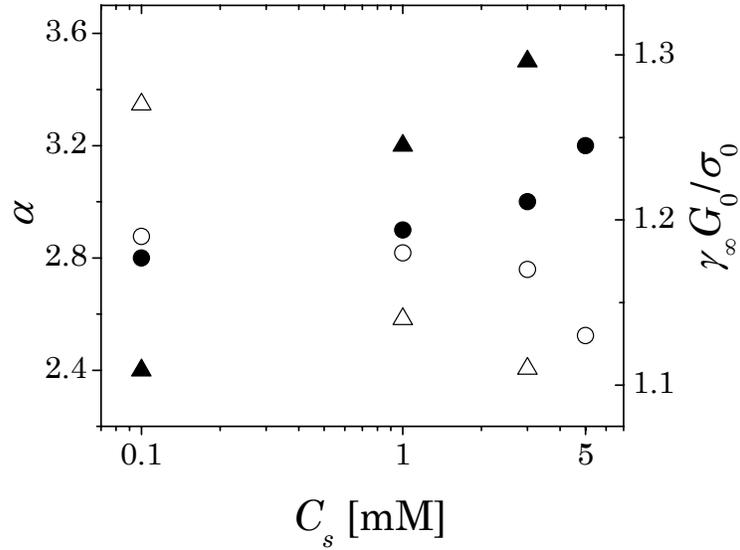

**Figure 2.** Comparison of $\alpha$ and normalized ultimate recovery $(\gamma_\infty G_0/\sigma_0)$ as a function of salt concentration $(C_s)$. The experimental data is from figure 3 of reference 8. Triangles and circles represent creep experiments carried out at 1.5 Pa on the sample having constant value of elastic modulus 191 Pa and 314 Pa respectively. Filled symbols represent $\alpha$ while open symbols represent normalized ultimate recovery.

## IV DISCUSSION

Reddy and Joshi[8] observed lesser recovery for suspensions with higher concentration of salt. In figure 2 we have plotted normalized ultimate recovery as a function of salt concentration. The corresponding value of parameter $\alpha$ estimated from equation 11 is also plotted on the same figure. It can be seen that increase in the concentration of salt lead to narrower distribution of relaxation modes (increase in $\alpha$). In addition, figure 2 also suggests that for the system having no salt, ageing lead to narrowing of relaxation time distribution while at higher salt concentrations ageing lead to broadening of relaxation time distribution [Elastic modulus of 191 Pa represents lower age while that of 314 Pa represents higher age]. In this section we discuss possible interrelation between the relaxation time distribution and the microstructure of laponite suspension. As discussed before, microstructural state of the aqueous suspension of laponite with varying concentration of salt is significantly debated in the literature.[13-15, 17, 18, 21, 22, 26-28, 32-34, 47, 48, 53-60] At the low ionic concentration (~ 0.1 mM) of salt, system is believed to be more homogeneous, though the precise nature of the microstructure is under debate. Various studies have proposed two different origins for nonergodicity, namely overall repulsive among particles[14, 18, 22, 26-28, 34] and attraction between a negative face and a weakly positive edge.[21, 47] At concentration of salt above 1 mM, the ergodicity breaking is believed to result in a fractal gel formation due to enhanced attraction between laponite particles due to screening of negative charges.[27, 28, 34] Nicolai and Cocard[29] observed the decrease in correlation length with reduction in concentration of salt. Correlation length is that length-scale of the system beyond which it is homogeneous. In view of this discussion, it appears that the homogeneity of the microstate decreases with increase in the concentration of salt.[21-23, 29, 47]

The trapped state of a laponite particle in the suspension, irrespective of the concentration of salt, can be represented by an energy well. The deeper the well, the larger is the cage diffusion time scale. Therefore the distribution of energy well depths also corresponds to distribution of relaxation modes. In the case of low ionic concentration of $Na^+$ ions, the trapped particle has predominantly repulsive interactions with its neighbors along with the attractive interaction between an edge and a face of adjacent particles. In addition, in this state, the length-scale at which the system is homogeneous is expected to be of the order of the particle length scale. With increase in the concentration of salt, the correlation length increases and consequently an individual particle has fewer nearest neighbors compared to that of a state at lower concentration of salt. Furthermore increase in the salt concentration leads to progressive screening of the negative charge on the laponite particle, so that the interactions among the particles are primarily attractive in origin. Assuming that the nearest neighbor interactions are the most dominating, it can be conjectured that at lower concentration of salt (progressively more homogeneous state), by virtue of greater number of nearest neighbors and repulsive as well as attractive interactions, distribution of barrier heights of energy wells is expected to be broader compared to a lesser homogeneous state, wherein an individual particle has fewer neighbors with primarily attractive interactions.[8] Since distribution of energy well depths has direct correlation with distribution of relaxation modes, it is plausible that increase in the concentration of salt lead to decrease in the ultimate recovery. Interestingly, our recent study on ageing under oscillatory stress field also suggested broadening of distribution of relaxation modes with decrease in the concentration of salt.[49]

Figure 2 also shows that ageing in the system with and without salt leads to narrowing and broadening of relaxation time distribution respectively. In our previous work we proposed that ageing in the system without salt leads to progressive ordering in the same, however such effect was not observed for the samples with salt.[34] Therefore, though we do not understand this behavior clearly, the different trends for systems with and without salt are not totally unexpected.

Finally, it should be noted that the age dependence of creep recovery behavior (and hence the relaxation time distribution) can be descried by equations 1 and 5. Equation 5 describes the relaxation time distribution $\tau_k$ given by $\tau_k \sim \theta_k^{1-\mu_k} t_w^{\mu_k}$. Here $\theta_k$ and $\mu_k$ are the model parameters. This equation is expected to give the distribution of relaxation time scales given that the dependence of $\theta_k$ and $\mu_k$ on age and stress is known. In equations 8 and 9, we are just approximating the distribution by Spriggs empirical relaxation spectrum. Taking into account the age dependence of $\tau_k$ does not give any significant advantage as dependence of $\mu_k$ on age and stress is not known. Therefore, considering the empiricism involved, we believe that this model is more suited for qualitative understanding of how distribution of relaxation times can influence ultimate recovery behavior in the limit of weak dependence of elastic modulus on frequency. Significantly, the model captures the observed trend of influence of relaxation time distribution on the ultimate recovery.

**V. SUMMARY:**

In this paper we develop an empirical model to correlate creep recovery behavior of ageing aqueous suspension of laponite to the relaxation time distribution of the same. We propose that in the limit of smaller creep-recovery time compared to waiting time (age), the time translational superposition can be applied to the ageing systems. In addition, the nonergodic systems are known to show weak dependence of elastic modulus on frequency. In this limit, multimode linear viscoelastic model and Spriggs relaxation spectrum[51] was solved to yield dependence of ultimate recovery on the relaxation time distribution. The model predicted that broadening of relaxation time distribution enhances ultimate recovery. We compare these results with the recovery behavior of aqueous suspension of laponite that shows less recovery with increase in the concentration of salt. This represents narrowing of the distribution of relaxation times with increase in the concentration of salt. We discuss this result in the context of microstructure and phase behavior of aqueous suspension of laponite.

**Acknowledgement:** This work was supported by DAE, BRNS young scientist research project.